\documentclass[USenglish,twocolumn]{article}
\usepackage{xcolor}
\usepackage[utf8]{inputenc}
\usepackage[big]{dgruyter}

\begin{document}

  \articletype{Research Article{\hfill}Open Access}

  \author*[1]{Bozena Czerny}


  \affil[1]{Center for Theoretical Physics, Polish Academy of Sciences, Al. Lotnikow 32/46, 02-668 Warsaw, Poland, E-mail: bcz@cft.edu.pl}


  \title{\huge Modelling broad emission lines in active galactic nuclei}

  \runningtitle{Modelling BLR in AGN}


  \begin{abstract}
{Broad Emission Lines are the most characteristic features of Active Galaxies, but the mechanism of creating a medium able to emit these intense lines is not quite clear. Observations clearly indicate that the motion of the material is predominantly Keplerian, with traces of inflow and clear signatures of outflow, but this still does not point out whether the lines partially come from the disk surface, or exclusively from the circumnuclear material, and whether this material originates from the disk as a wind, or comes, at least partially, from outside.  I review the basic scenarios for the formation of the Broad Line Region (BLR), and the recent progress in modelling the physical conditions in he emitting medium. The current state is the outer radius of the BLR is fixed by the dust sublimation temperature in the medium exposed to the irradiation from the central source, the inner radius is likely fixed by the dust sublimation temperature in the atmosphere of the non-illuminated accretion disk, and the local cloud density is a universal number fixed by the radiation pressure confinement. The time-dependent aspects of the BLR formation, however, still wait for serious modelling effort matching the quality of the observational data.}
\end{abstract}
  \keywords{Active Galactic Nuclei, Emission Lines}

  \journalname{Open Astronomy}
\DOI{DOI}
  \startpage{1}
  \received{..}
  \revised{..}
  \accepted{..}

  \journalyear{2014}
  \journalvolume{1}

\maketitle
\section{Introduction}

Carl Seyfert selected six galaxies (or extragalactic nebulae, as he refereed to them) for a detail studies because of their bright and compact nuclei and puzzling broad and strong emission lines \citep{seyfert1943}. These objects are still with us: NGC 4151, NGC 3516, NGC 7469, NGC 1068 and NGC 1275 (least popular but gaining recently more attention since it is located at the center of the Perseus cluster observed by Hitomi X-ray satellite). With this paper he pioneered the systematic study of the Broad Line Region in Active Galactic Nuclei.

Since this period our general understanding of AGN improved considerably. AGN emit radiation of predominantly non-stellar origin, although compact nuclear cluster is also an important part of the system. This non-stellar emission comes from the material surrounding the central supermassive black hole, and the source of the energy is the dissipation of the energy accreting onto black holes. Supermassive black holes exist in all regular galaxies, including the Milky Way galaxy, but the central activity depends on the amount of material which at a given epoch is available for accretion. Milky Way is currently so weakly active, and this activity can be observed only because of the proximity of its nucleus. Milky Way galaxy has been more active in the past, as we know from the light echo of the past activity, and there are many arguments that the activity of galaxies has the intermittent character. Nevertheless, since in general the characteristic timescales for major changes in AGN are at least hundreds or thousands of years (we will return to this point in Sect.~\ref{sect:CL}), galaxies are classified according to their actual activity level. However, in the short timescale AGN also vary which helps to understand their structure when multi-band monitoring studies are performed. This variability is of stochastic nature, the fastest variations are seen in the X-ray band, which implies that the X-ray emitting region is the most compact one.

A fraction of AGN emit strong and broad emission lines, with the full with half maximum (FWHM) from over 1000 km s$^{-1}$ to some 15 000 km s$^{-1}$, or even 27 000 km s$^{-1}$ measured as full width zero intensity (FWZI) in a source recently studied with HST \citep{bianchi2019}. These broad lines are of key importance for the study of galaxy evolution and for potential cosmological application because they allow for measurement of the central black hole mass in AGN surveys. Methods use the assumption of the predominantly Keplerian motion of the material being the line source, and the distance from the black hole is measured from the time delay of the emission line with respect to the varying continuum. Such a delay has been measured now in over 100 sources and it allowed to find the relation connecting this radius to the absolute source luminosity for a known cosmology. Such relation, combined with the measurement of the line width, allows to calculate the black hole mass from a single spectrum. BLR offer prospects to measure the absolute luminosity of the source in a redshift-independent way either by identification of sources emitting close to the Eddington ratio or by combining the measured size and the theoretical R-L scaling, or observational scaling obtained for nearby sources. The BLR geometry is clearly flattened, concentrated relatively close to the equatorial plane, since BLR material is rarely seen in absorption. This picture of the flattened distribution of clouds roughly in Keplerian motion had been recently confirmed by the first resolved image of the BLR obtained in the IR in the Paschen$\alpha$ line with the GRAVITY instrument \citep{GRAVITY2018}. 

The progress in understanding of the energy source in AGN is not matched, however, by the current level of understanding of how BLR actually forms. Some attempts on parametric modelling are done since many years but the issue is complex. As in the case of the early studies by Carl Seyfert, here the observational studies of the line variability are well ahead of the theory. The key information comes from observing of a single source systematically for many years, and here Alla I. Shapovalova's enormous contribution comes into play. She authored/co-authored over one hundred papers on how the line profiles in selected AGN change. We have to learn how to exploit fully such data, not just for time delay measurement, but to understand what is the origin of the emitting material, where it goes (escapes ? returns to accretion disk ?), what it tells us about the accretion flow close to the black hole and the emitted radiation (BLR clouds do not necessarily see the same continuum as we do!).

In this review we will concentrate on galaxies which are (i) strongly active (ii) their nucleus is not shielding from the line of sight by the dusty/molecular torus (iii) their optical/UV spectra show broad emission lines. In such galaxies the broad band spectral energy distribution (SED) is dominated by the optical/UV/soft X-ray Big Blue Bump which comes from the optically thick geometrically thin accretion disk surrounding a black hole. Close to the black hole a very hot plasma exists, frequently referred to as a hot corona. The geometry of this component is still under discussion but the element is required to explain the hard X-ray emission which carries a fraction ($\sim$ 5 to 50 \%) of the total energy. Some of these sources are also radio-loud, i.e. they have a strong jet component which dominates in the radio and gamma-ray band. We do not discuss blazars in this review since their Doppler-boosted jet directed towards observer make the study of the BLR rather difficult.  

I review the basic scenarios for the formation of the BLR, and the recent progress in modelling the physical conditions in he emitting medium. Sect.~\ref{sec_obs} gives a brief introduction to the observational progress, Sect.~\ref{sec:mass} introduces the issues related to black hole mass measurements. Parametric approach to BLR modelling is described in Sect.~\ref{sec:param}, and the physical conditions in the BLR summarized in Sect.~\ref{sec:phys}. The theories of the origin of BLR are outlined in Sect.~\ref{sec:orig}, and Sect.~\ref{sect:wind} is devoted to more detail discussion of accretion disk winds. Relatively unexplored observations of time-dependent line profiles and the Changing Look AGN are briefly presented in Sects.~\ref{sect:profiles} and \ref{sect:CL}. Finally, conclusions are comments on the future progress are given in Sect.~\ref{sec:conc}.

\section{Observational background}
\label{sec_obs}

BLR is a very complex region. The emitting material is frequently referred to as 'clouds', and the modelling faces all the issues of a weather forecast. Thus the progress in the understanding of BLR structure relies mostly on observational progress, with some help from the physically based studies of ionization state of the material, line emissivity, radiative transfer, dynamics related to radiation pressure acceleration etc.

Early studies of the BLR concentrated on the photoionization equilibrium and the comparison of the modelled line ratios with the observed one \citep[e.g.][]{osterbrock1978}. This is not simple since the temperature of the clouds range from 10 000 K to 20 000 K for a broad range of the ionization parameter, U \citep[see][for a review]{peterson2006}.  Also the line profiles were used with the aim to get constraints for the BLR dynamics but the results were not conclusive \citep[e.g.][]{capriotti1980}.

The essential progress came with the development of the reverberation mapping approach. AGN continuum is variable, and delayed response from the BLR allows for relatively direct insight into its location, extension and structure \citep{peterson1998}. Early monitorings were done mostly in the optical band \citep[e.g.][]{lyuty1977,kaspi2000,peterson2004,bentz2013}, but also in UV using IUE satellite \citep[e.g.][]{clavel1991}. \citet{wandel1999} showed BLR size measurement from reverberation mapping gave much more precise measurements than estimates based on ionization parameter, supported the view that BLR is predominantly in Keplerian motion and can be used to measure the black hole mass. 

The basic technique of the time delay measurement is simple. The continuum lightcurve and the emission line lightcurve have to obtained from spectroscopic observations. In the case of H$\beta$, the continuum is conveniently measured at 5100 \AA, and the spectra are usually calibrated using the [OIII]5007 line which form in the NLR. Next, time delay has to be measured, using for example the Interpolated Cross Correlation Function (ICCF; citealt{peterson1998}), or alternatively, more recently developed JAVELIN software \citep{zu2011}, based on the assumption that AGN lightcurves are well represented by Damped Random Walk \citep{kelly2009}. Other methods, like ZDCF or $\chi^2$ method can also be used (see, for example the test ov various methods for a single source in \citealt{czerny2019}). In practise, the time delay determination has to be done carefully, the stellar contamination should be removed to get proper radius-luminosity relation \citep[e.g.][]{bentz2013}, the normalization of the spectrum has to be done at the basis of calibration stars if the UV spectral range is probed, or the Eddington ratio is high and [OIII]5007 line is weak and strongly contaminated by Fe II emission. 

Recent monitorings were considerably expanded in several directions. New parameter space started to be studied, with reverberation concentrating on high Eddington ratio sources \citep{dupu2014,dupu2015,dupu2016,dupu2018}, and on distant quasars \citep{lira2018,czerny2019,grier2019}. Second, higher quality reveration starts to allow for wavelength-resolved delays which allow for real mapping of the BLR \citep[e.g.][]{grier2013,dupu_resolved2018,derosa2018}. Finally, instead of studying individual sources, a whole field of view is monitored, thus allowing for the measurement of time delays for several objects \citep{king2015,grier2017,grier2019}. There is also a tremendous effort in the opposite direction: study a single object, like NGC 5548, but putting on an impressive dense in time multiwavelength campaign showing the source behaviour from the IR through optical and UV to X-rays, aimed mostly at understanding the continuum delays and multi-band entanglement \citep[e.g.][and the references therein]{derosa2015,kriss2019}, but even in this case the BLR reprocessing apparently plays a role \citep{cackett2018}, and the discovered anomaly of the UV line behaviour at some period of time \citep{goad2016} called a lot attention. 

In addition to this line of research, mostly focused on time delays, there is a long albeit not very extensive trend to monitor the variations of the line shape in individual objects. A number of papers on this issue was authored/co-authored by Alla Shapovalova,, as listed in \ref{sect:profiles}. This long-term effort still waits for better exploitation by teoreticians, but it gains importance in view of the two fashionable aspects: possibility of existence of close pairs of binary black holes (important in broad context of gravitational waves) and increasingly well documented rapid state transitions in Changing-Look AGN.

\section{Virial factor and the black hole mass measurement}
\label{sec:mass}

Modelling the BLR is important for the overall understanding of an active nucleus but it is also a key element of the black hole mass measurement, later broadly used in cosmology. The basic principle is simple: we use the Kepler law 
\begin{equation}
    M = f {R_{BLR} v^2 \over G},
\end{equation}
so we need to measure the velocity $v$ and the radius $R_{BLR}$. As a proxi for velocity, the line width is used, either FWHM, or the line dispersion. The radius is measured directly from the time delay, or from the radius-luminosity scaling. The problem is with the scale factor, or virial factor, $f$. It contains all uncertainties related to the viewing angle, BLR extension, cloud orbits etc. If the cloud distribution is spherically symmetric, and FWHM is used, $f = 3/4$ \citep[see e.g.][]{kaspi2000}. However, there are no doubts that the distribution is flattened, and the BLR is extended, so the radius in the formula is a kind of effective radius for a given emission line. Thus, the virial factor has to be calculated. It is also expected that the factor can depend on the viewing angle, and perhaps on the properties of the AGN class (e.g. Eddington ratio).

Virial factor can be estimated observationally, if an independent method of the black hole mass measurement is used. In this way we generally obtain a universal inclination-dependent value \citep[e.g.][]{peterson2004}. Recent attempt include the effect of the viewing angle by finding a statistical relation between the virial factor and the line width, using an independent black hole mass determination based on accretion disk continuum fitting \citep{mejia2018}. 

However, the ultimate goal is to be able to dismiss the virial factor and to calculate the black hole mass directly from the model, knowing the 3-D BLR structure. 

\section{Parametric approach to the dynamics of the  BLR}

\label{sec:param}

Full model of the BLR should consist of the following elements (i) physically motivated dynamics (ii) radiative transfer computations of the region emissivity (iii) computations of the line profiles as a function of the viewing angle toward an observer, ready to be compared with the observational data. This requires 3-D time-dependent modelling, and such a complexity is till beyond the current possibilities. The weakest point is actually the dynamics, while there are very advanced codes capable of doing stationary radiative transfer (e.g. CLOUDY, \citealt{CLOUDY2017}). Therefore, considerable effort was done using a parametric approach to the  dynamics of the BLR. This line of the study brought significant achievements and is extremely practical in analysis of the observational data.

In parametric approach to the dynamics, the distribution of the clouds is set by a number of parameters. First, an assumption has to be made with respect to the location of an accretion disk with respect to the BLR. In most models, the accretion disk is assumed to have the outer radius smaller than the inner radius of the cloud distribution, so clouds can move on the inclined orbits crossing the equatorial plane. Alternatively, if the accretion disk is still present in the BLR, such orbits are not possible because the clouds cannot cross the dense disk. In this case only parts of the Keplerian orbits are available. 

The BLR range is usually constrained by the adopted inner and outer radius of the zone. Then, the number of clouds as a function of radius and distance from the equatorial plane has to be specified. Next, the individual cloud properties have to be assumed. In most models, the local density of the clouds is a decreasing function of the radius. Also the total cloud size, usually given in the form of the total hydrogen column density is specified. Most advance models use the locally optimized cloud (LOC) approach \citep{baldwin1995}, where at a given radius clouds of different densities can be located, and the number of clouds as functions of density and radius are specified through a power law dependence, with power law indices as model parameters.  Most computations were performed using constant density approximation inside a cloud, although recently more models turn toward constant pressure approach. Clouds should be approximately in thermal equilibrium to last longer than the dynamical time, so they should be denser on the colder size and less dense on the illuminated, hotter part. On the other hand, if the optical depth of a cloud is small, or the local density of the cloud is high, the difference between the constant density and constant pressure cloud structure is not very large.

Having cloud distribution, one can compute the cloud emissivity with the radiative transfer codes like CLOUDY, and then study the line ratios, depending on the shape of the incident radiation, clod metallicity etc. In most advanced models, the emissivity is calculated taking into account the light travel time and the variable incident continuum, and having the cloud motion allows also to calculate the line profiles from the model. Such models have additional advantage that they do not need any assumption about the virial factor. 

\subsection{Outer radius of the BLR}

One of the important achievements from this line of modelling was the determination of the outer radius of the BLR. Observationally, many AGN show a complex H$\beta$ line profiles, with a narrow peak at the line center superimposed on the broad component. This observational fact motivated the division of the emission lines into coming from the narrow line region (NLR) and BLR, and the clear distinction suggested a physical gap between the two regions, resulting is well separated velocity range. The mechanism of this separation has been found by 
\citet{netzer1993}. They assumed that the material forming the BLR extends in a continuous way all the way from the inner BLR out to NLR. They assumed a power law dependence of the single cloud density and total column density on the radius and they calculated the emissivity profiles for a number of emission lines. What was important, they included the fact, that at some point the distance from the nucleus was large enough to allow for the existence of the dust. At dust sublimation radius the emissivity of all lines dropped significantly, by orders of magnitude, since the dust efficiently intercepted the ionizing photons, and the line emissivity only recovered at very large distances. Thus, they showed that the dust sublimation radius sets the outer radius of the BLR.

However, not all sources show such a clear separation between the NLR and BLR. Highly accretion sources like narrow line Seyfert 1 (NLS1) galaxies, with line width smaller than 2000 km s$^{-1}$ and more massive, type A quasars, with line width below 4000 km s$^{-1}$ are frequently well modeled with a single Lorentzian profile (\citealt{veron_cetty2001,sulentic2002,zamfir2010}; see also \citealt{goad2012} for the interpretation). In \citet{adhikari2016} (see also \citet{adhikari2018} for constant pressure models) we applied the original model of the \citet{netzer1993} to these objects. The use of appropriate SED was not enough to explain the phenomenon, but when we used much higher local cloud density than in \citet{netzer1993} (i.e.  $10^{11.5}$ cm$^{-3}$  instead of $10^{9.4}$ cm$^{-3}$ at the sublimation radius) we observed no gap between BLR and NLR at the dust onset. Only Fe II emission still showed significant drop there.

\subsection{Application to velocity-resolved maps and inflow/outflow tests}

High quality reverberation mapping brings not only the measurement of a single number - time delay of the line emission with respect to the continuum but the line transfer function (response of the line emission to delta-function impulse from the central source) or even the whole velocity-resolved delay maps \citep{grier2013,derosa2018}. To read such maps, simple basic models of the dynamics were tested (Keplerian motion of a ring, inflow, outflow). The models developed for that purpose are relatively simple stationary models of the BLR geometry and velocity field which allow for construction of the time-delay maps for qualitative comparison. Unfortunately, such measurements require long and dense monitoring, and were done only for a few sources. Also errors in the maps are rather large, so direct map fitting between the model and the data are not yet attempted.

\subsection{Application to the advanced time-delay measurements}

Much more careful comparison between the models and the data is done if the measurements concentrate only on time delays and the mean shape of the line instead of whole velocity-resolved map. Such results are much more interesting that just the simple time-delay measurement as it brings also the insight into the region structure.

Parametric models are very flexible so they are very convenient for that purpose. Fitting the data is not simple since the number of free parameters is the model is considerable so some Monte Carlo approach is in general necessary. But the results are excellent: model can reproduce the continuum lightcurve, the selected line lightcurve, and the mean line profile \citep{grier_mod2017,pancoast2018,li2018}. Such modelling shows that a considerable complexity of the BLR is necessary, and a two-zone model is needed for higher Eddington ratio sources to explain the time delay relatively shorter (for a given monochromatic luminosity) than in lower Eddington sources \citep{bentz2013}.

These models strongly support the basic facts about the BLR which accumulated over the years:
\begin{itemize}
\item the BLR material is roughly in Keplerian motion;
\item with some signatures of outflow, stronger in case of
high ionization lines;
\item some(times) signatures of inflow;
\item the distribution is flattened, but the covering factor is
high (0.1 - 0.3) so part of the material is far from the
equatorial plane.
\end{itemize}

\section{Physical conditions in the BLR}
\label{sec:phys}

Parametric models discussed in the previous section create a complex picture of the BLR. In addition, theoretically we also expect that at least four important parameters should describe an active nucleus: black hole mass, accretion rate, spin and the viewing angle. On the other hand, local BLR properties which we would like to measure are the ranges of temperatures, densities, local velocity (apart from the overall Kelerian motion), turbulent velocity, radiation field within BLR and sources of heating. From the reverberation mapping it is quite clear that radiative heating dominated, but contribution from shock heating, for example, might be also important. But what me measure, are line intensities and line width, so the connection of this rich information to the physical parameters and global source parameters is not simple and direct. Formal principal component analysis (PCA) implied that some of these parameters are relatively unimportant and multiple measured quantities like line ratios, line widths, absolute luminosity and broad-band spectral indices are correlated, and the sources form a quasar main sequence \citep{boroson1992,sulentic2000,marziani2018}. The goal to pass uniquely from the modelling to the observed trends is not easy to achieve \citep{panda2018,panda2019}.

As mentioned already in Sect.~ \ref{sec_obs}, the only parameter which is well constrained is the temperature, in the range of 10 000 - 20 000 K, which comes from the efficient heating/cooling of the partially ionized plasma due to atomic transitions keeping the temperature almost independently from the ionizing flux (see e.g. \citealt{peterson2006}). This stable temperature range is well visible in the study of thermal instability in the irradiated plasma \citep{krolik1981}; only very intense irradiation by hard ionizing flux, as measured with respect to the local density, leads to almost complete or complete ionization and the heating/cooling balance is then based on Compton heating/cooling, driving the temperature up to $\sim 10^7$ K range in unstable fashion. However, if indeed we want to obtain precise measurement of the BLR temperature the goal becomes complex since there are no particularly suitable line ratios to use for that purpose \citep[see e.g.][]{ilic2008}.

Estimates of the density are more complex and the density values in principle can depend on the source. Recently the quasar main sequence was successfully modelled with the code CLOUDY introducing correlations between the Eddington ratio, cloud density, metallicity, viewing angle and line width \citep{panda2019}. Some of this correlations are likely caused by AGN evolution which should be further studied. 

Requested cloud local densities are rather high, close to $10^{11}$ cm$^{-3}$ at larger values of the Eddington ratio. However, this is nicely explained within the radiation pressure confinement scenario \citep{baskin2018} where this number appears a universal condition balancing the radiation pressure from the central source, independently from the central source parameters. 

The important aspect of the BLR regions is its dynamics, apart from the Keplerian motion, which is neglected in the simple estimates of the local density and temperature. As indicated by line intensity modelling, the medium is turbulent, with the turbulent velocities of order of 10 - 20 km s$^{-1}$ in the Fe II formation region \citep[e.g.]{bruhweiler2008,panda2018}, and likely higher closer in, so the turbulence is formally supersonic. Observations of LIL part of BLR clearly indicates outflow, with velocities in CIV lines up to a $\sim$ 1000 km s$^{-1}$, i.e. significant fraction of the local Keplerian speed \citep[e.g.][]{brotherton1994,marziani2017}. LIL part of the BLR carry less outflow signatures although line distortions in H$\beta$ are also observed \citep[e.g.][]{boroson1992,dupu2018}. The outflow can carry a lot of mass although the estimates of the outflow rate is rather difficult \cite[e.g.][]{shin2017}. Massive outflows seen as Broad Absorption Line may form at larger distances, not within the BLR \citep[e.g.][]{moe2009,hamman2019}. Also the connection of the Ultra Fast Outflows (UFO) with the BLR is not clear, but not excluded, and these highly ionized outflows also can carry significant mass \citep[e.g.][]{tombesi2010}.

The missed element of the modeling is certainly the level of clumpiness in the flow. Smooth profiles of the emission lines in many sources imply large number of the clouds or a continuous medium, while (rare) X-ray obscuration events imply well formed clouds with dense head with the hydrogen column density of order of $10^{23}$ cm$^{-2}$ and a cometary tail \citep[see e.g.][]{bianchi2012}. It is quite likely that, like in stellar winds, the outflow starts as a uniform flow, and clumpiness develops later on due to the know thermal instability in a medium illuminated by relatively hard central source \citep{krolik1981}. The details of this process must depend on the actual scenario of the BLR formation. This seems more plausible that the development of a mist of $10^{17}$ cm$^{-2}$ cloudlets advocated by \citet{mccourt2018}. 

\section{The models of the BLR origin}
\label{sec:orig}

Parametric approach gives some hint about the dynamics but will never tell us directly why this material is there. For that, we need a separate approach, which at this moment is not yet very advanced, but it is very important for the deeper understanding of the role of the BLR in the overall AGN activity phenomenon.

The main ideas about the origin of the BLR material can be divided into four main scenarios: (i) inflow models, where the BLR material arrives from the outer regions (ii) formation 'in situ' from fragmentation of the accretion disk (iii) irradiated disk surface (iv) accretion disk wind. Below we discus shortly these main scenarios, and later in detail the issue of the disk wind since this scenario is relatively well developed and has several different variants.

\subsection{Inflow models}

It is a paradox that AGN are powered by accretion but in observations, when there is a possibility to measure directly the velocity field of the matter we predominantly see the outflow (jet, winds). This is because the velocity in the accretion disk is very small in comparison with the local Keplerian orbital velocity. However, velocity-resolved maps suggest that in some sources a net inflow takes place \citep{}, and there is a claim of the inflow seen from the shift between the Fe II emission and the AGN rest frame \citep{hu2008}. This can explain the high Fe II intensity since we see mostly the shielded (non-illuminated) sides of the infalling clouds \citep{ferland2009}. On a theoretical ground, active nucleus is surrounded by nuclear stellar cluster, frequently a dusty/molecular ring, and dusty/molecular torus. The feeding scenarios of AGN also imply formation of gas streams as a result of the gravitational instabilities in the galaxy central region, including 'bar within bar' scenario \citep{shlosman1989}. This inflowing material can be fragmented into clouds, fall towards the center due to possessing low angular momentum, and finally forming a ring-type structures at a circularization radius. Such a scenario was considered in detail by \citet{wang2017}. 

An interesting option has been discussed by \citet{elvis2017} where the author considers the outflow from the disk, subsequent cloud formation due to the thermal instability \citep{krolik1981}, and the fallback of the newly formed heavy clouds not supported strongly enough any more by the radiation pressure.

\subsection{Formation 'in situ'}

Standard accretion disks surrounding a supermassive black hole become gravitationally unstable at large radii. For some intermediate radii the disk can become marginally stable due to the development of turbulence and spiral waves transporting angular momentum, but finally at still larger radii, gravitational instability wins (unless the disk is strongly dominated by the magnetic field). Gravitational instability leads then to disk fragmentation and star formation. Detailed model of the complex processes taking place there was already developed by \citet{collin_zahn1999,collin_zahn2008}, and later studied by \citet{wang2011}. Supernovae outbursts taking place there can eject copious material thus forming BLR clouds, and also provide enrichment of the gas in heavier elements. It also can lead to repeating episodes of the BLR formation, as proposed by \citet{wang2012}. Comparison of predictions from this mechanism with the observed location of the BLR did not show agreement \citep{czerny2016} but it was not based on newer results of reverberation measurements which imply considerably shorter delays, particularly for higher Eddington ratio sources \citep[][]{dupu2015,dupu2016,grier_obs2017,dupu2018,maryloli2019}.

\subsection{Irradiated disk surface}

Broad emission lines can be further divided into high ionization lines (HIL), like He II or CIV, and low ionization lines (LIL), like H$\beta$, Mg II or Fe II \citep{collin1988}. HIL must originate in some form of a wind since they show signatures of outflow, while LIL are more consistent with purely Keplerian motion and originate close to the disk. It can actually, at least partially, from the disk surface. Direct emission from the disk surface seems suitable for double-peak LIL line profiles \citep[e.g.][]{eracleous2003,bianchi2019}. As discussed by \citet{kollatschny2013}, in general case one needs a combination of a Keplerian velocity and a turbulent velocity, but apparently this turbulent velocity component rises in importance with the Eddington ratio of the source. Such trend is consistent with expectations of the dust-based BLR model to be discussed in the next section. Double-peak line sources have in general low values of the Eddington ratio, consistently with this trend. Recently, the static irradiated disk atmosphere, puffed up by in the region dominated by the dust opacity, was proposed as a BLR model by \citet{baskin2018}, independently from the Eddington ratio of the source but the authors did not show yet the predicted line profiles from their model.

\subsection{Accretion disk wind}

As mentioned before, this is the most complex scenario. The theoretical requirement common to all the variants is that the underlying stationary cold optically thick disk extend as far as the BLR region in all sources. The advantage of the model even in its most general version is that it predicts absence of the lines in sources which do not show the presence of the Big Blue Bump. At low valued of the Eddington ratio (and in true Seyfert 2 galaxies, where the broad lines are intrinsically not present instead of just being shielded from our line of sight), in general, we do not expect the cold accretion disk and the accretion flow proceeds in the form of a hot flow (e.g. advection-dominated accretion flow - ADAF; see \citealt{feng2014} for a recent review). This is the case of Milky Way, as well as M87 recently mapped by Event Horizon Telescope \citep{EHT2019}. Hot inflow can be also accompanied by the wind, but such wind is fully ionized and not capable of producing BLR lines. This is possibly characteristic for accretion in elliptical galaxies. However, the paper by \citet{bianchi2019} teaches us that some 'apparent' true Seyfert 2 galaxies can have cold disks and BLR, if properly observed. 

The advantage of the scenario is that actually the presence of the winds from accretion disks in AGN is expected \citep[for a review, see][]{proga2007}. Specific predictions depend on the assumption about the outflow driving force and they are discussed in more detail in the next section.

\section{Accretion disk wind mechanisms and their consequences for the BLR formation}
\label{sect:wind}

The predictions of the BLR model based on the accretion disk wind idea depend on the mechanism driving the outflow. This could be magnetically (centrifugally) driven wind, thermally-driven winds, line-driven wind, or dust-driven wind. Additionally, some shielding is required in some of these scenarios. Some of these models are numerically very advanced, combining the hydrodynamical simulations with the radiative transfer. Some of the models predict outflow, while the others produce failed wind, with the material returning again to the disk.

\subsection{Magnetically driven winds}

If the accretion disk is permeated by the large-scale magnetic field, the magneto-centrifugal force will drive the outflow \citep{BP1982}. The drawback of the model is that it requires the existence of the strong large-scale poloidal component which cannot be produced inside the disk by magnetorotational instability \cite{balbus1991} but instead must be driven in by the accretion flow from the interstellar medium. Also the transonic part of the wind launching is not that simple in this model \citep{proga2007} but most applications are based on self-similar solutions easy to apply which guarantee that the wind has the expected properties \citep[e.g.][]{everett2005}. The amount of outflowing material cannot be predicted by this model but the interesting firm prediction of the model is the concave shape of the flow lines, which in principle can be effectively used to test models if this wind is also occasionally responsible for absorption lines. 

Emission line profiles were not calculated from these models, but the flow, in the dusty part, can well represent the dusty molecular torus \citep{elitzur2006,keating2012}. On the other hand, \citet{fukumura2015} proposed that such winds capable of achieving high velocities and high ionization states can account for the Ultra Fast Outflows (UFO) seen as absorption features in X-ray data. The launching radius in this case should be of order of $\sim 200$ R$_g$. 

\subsection{Thermally-driven winds}

If the temperature of the accretion disk outer layers approaches the virial temperature spontaneous outflow from the disk will form. This may happen for example in the outer disk parts if the disk is irradiated by central source to the Inverse Compton temperature \citep{begelman1983}, if there is an energy dissipation close to the disk surface \citep{czerny1987}, in the presence of a hot corona above the disk \citep{witt1997}, or in the case of the inner hot flow \citep{blandford1999}. This outflow is likely hot, and in most cases is not expected to provide directly emission lines. However, \citet{krongold2010} and \citet{mizumoto2019} proposed that thermally driven winds from the outer BLR or from the dusty torus region can be efficient enough to explain the warm absorber phenomenon in AGN. 

\subsection{Radiation-pressure driven wind}

Here a next level sub-division should be done since two sources of opacity are under consideration. Line-driven wind is powered by absorption in lines, and the force multiplier (the ratio of the actual radiation pressure to the radiation pressure expected from Thompson scattering) can achieve very high values, or order of a few hundred, for optimum level of material ionization.
But also dust scattering is an efficient source of opacity and can drive the outflow when the irradiation is low enough to keep the dust temperature below the sublimation temperature. We know from the properties of stellar winds that line-driven winds (e.g. in O-type stars) are faster and carry less material, while dust-driven wind (e.g. in AGB stars) are much slower but carry more mass. 

\subsubsection{Line-driven wind}

Line-driven winds as a formation mechanism was studied in detail already by \citet{murray1995}. \citet{chiang1996} and \citet{murray1997} showed that the wind model can produce not only double-peak profiles but also single-peak profiles. They also predicted the complexity of the line profiles, including the asymmetry between the red and a blue wing due to the combined effect of the wind velocity field and radiative transfer. The model adopted some simplifications in the description of the velocity field but line profiles were calculated and actually compared to the observational data for NGC 5548.
Even the response functions were provided for various line components. The model, according to the authors, required further improvement before it was ready to be used for data fitting \citep{chiang1996} but actually not much progress have been made along the suggested lines, and instead a simple to use parametric models are used for that purpose (see Sect.~\ref{sec:param}).

Instead, since there was a considerable progress in hydrodynamical simulations of the AGN winds, and \citet{waters2016} used their hydro models to predict the line shape and the transfer function for different viewing angles. In this approach the dynamics and the ionization state were not computed at the same time, the line transfer has been done a posteriori due to the model complexity. 

The importance of such studies is clearly shown by observations which imply that a significant fraction of AGN is capable of producing strong outflows, manifesting themselves in blueshifted or skewed [OIII] lines, and such winds are expected to deposit energy and momentum into the interstellar medium, thus regulating both star formation and supermassive black hole growth \citep[see e.g.][and the references therein]{vietri2018}.

\subsubsection{Dust-driven wind}

New approach to modelling of the BLR (or, more precisely, to its LIL part) has been proposed by \citet{ch2011} in which the launch of the outflow is caused by the radiation pressure acting on dust. Since BLR is closer to the black hole than location of the dusty torus, launching is possible close to a disk surface, where the irradiation by the central regions is still unimportant. At larger heights the material is irradiated, the dust evaporates, the radiation pressure force turns off, and the material falls back toward the disk. This failed radiatively accelerated dusty outflow (FRADO) well explains the location of the inner radius of the BLR - the BLR starts at a radius where the disk effective temperature becomes lower than the dust sublimation temperature for non-illuminated accretion disk. This radius in a standard accretion disks scales with the monochromatic flux, but does not depend separately either on the accretion rate, or on black hole mass and thus explains quantitatively the radius-luminosity relation \citep{bentz2013}. 

Preliminary results show that the model is promising \citep{czerny2015,czerny2017} but the line profiles must be calculated more carefully using 3-D cloud motion instead of 1-D approximation. In addition, new reverberation measurements show considerable departures from the standard radius-luminosity relation which in general systematically rise with the increase of the Eddington ratio \citep{maryloli2019}. It remains to be seen if 3-D FRADO model can explain this trend. The effect may be caused either by stronger shielding of the launching region or by anisotropic emission of the puffed-up accretion disk \citep{wang2014b,li2018}.  

The important difference between FRADO and the model of \citet{baskin2018} is that the second model is static in the vertical direction. FRADO outflow is based on the assumption that the opacity in the upper atmosphere of the disk, determined as the Planck mean is higher that the Rosseland mean opacity applicable in the disk interior, so there is a gap between the disk atmosphere in hydrostatic equilibrium and the surface where gravity balances the radiation pressure \citep{czerny2017}. Realistic estimates of the radiation pressure with the use of  KOSMA -$\tau$ code \citep{rollig2013} show that the equilibrium surface is indeed much higher than the disk surface (M. Naddaf, in preparation) calculated with all the opacity elements included \cite{czerny2016}. 

\section{Modelling time-dependent profiles}
\label{sect:profiles}

Searches for a shape changes in AGN is a very difficult task since the changes are generally slow, over timescales of years, so they are difficult to organize and pursue, but prof. Alla I. Shapovalova and her collaborators were able to purse that project for a number of sources (for a summary of the monitoring see e.g. \citealt{ilic2017}).

Monitoring of selected AGN over many years clearly show that the line profiles in some sources are changing significantly in time (3C 390.3 monitored for 12 years; \citealt{Sh2001,Sh2010b}; NGC 5548 observed for 6 years, \citealt{Sh2004}; NGC 4151 monitored for 11 years, \citealt{Sh2010}; Arp 202B observed for 23 years, \citealt{Sh2013}) while in other sources the line shape does not change much (Ark 564 observed for 11 years, \citealt{Sh2012}; NGC 7469 observed for 20 years, \citealt{Sh2017}; E1821+643 observed for 24 years, \citealt{Sh2016}; NGC 3516 observed for 22 years, \citealt{Sh2019}; in this last source, the amplitude of the line changes strongly, down to line disappearance). Similar monitoring, but not as long, has been performed by a few other groups (e.g. NGC 1097 \citealt{storchi1995}; 3C390.3, \citealt{vie1991}). 

This impressive data collection is not yet appreciated enough since models of such evolutionary changes are not fully ready. The list of likely explanations of the line profile variability contains: elliptical accretion disk, spiral waves inside a disk, or a binary black hole, particularly in the context when the asymmetry of the line changes semi-periodically. Binary black hole is particularly attractive after the discovery of gravitational waves. The likely effect of the secondary black hole on the BLR line shape was modelled by a number of authors \citet[e.g.][and the references therein]{gaskell1983,popovic2012,ilic2014,wang_bin2018}. Particularly powerful possibilities emerge if the GRAVITY instrument can be used to spatially resolve the source \citep{songsheng2019}.  The binary black hole mechanism was claimed to be responsible for the quasi-periodic changes in a number of well known AGN, e.g. in Ark 120 \citep{li_bin2019}, NGC 4151 \citep{bon2012}, NGC 7674 \citep{kharb2017}, NGC 5548 \citep{li_bin2016}. Line drifts in quasar spectra also suggested to be related to the presence of a binary black hole \citep[e.g.][and the references therein]{liu_bin2019} .

However, current models are not precise enough to differentiate between the proposed mechanisms. Disk precession gives rather similar predictions to the effect of the secondary, and differentiating between the two effects would have to be likely based on realistic estimates of the precession period in a given system and estimated of the dynamical perturbations of the secondary to the BLR region. The best case so far of a binary black hole is for the famous system OJ 287 \citep{valtaoja2000,lankeswar2019} but this source does not have a BLR, and but even in this case a possibility to consider a precession model without the presence of the secondary black hole exists \citep{britzen2018}.      

\section{Changing-Look AGN}
\label{sect:CL}

AGN always vary in the optical/UV band (as well as in other energy bands) but usually the variability has a stochastic character and the variability amplitude is moderate. However, extreme example of the AGN variability occasionally happen. In some cases the BLR forms suddenly in a galaxy which did not seem active before, in some other well known AGN lines temporarily almost disappear which leads to a change of the source classification, in extreme cases almost from Seyfert 1 to Seyfert 2 \citep[e.g.][]{cohen1986,storchi1993,Sh2004,lamassa2015,bon2016,mcelroy2016,macleod2019,oknyansky2019,trakhtenbrot2019,trak2_2019,katebi2019,Sh2019}. A name name was invented for these sources - changing look (CL) AGN. The phenomenon was occasionally observed before but now the number of identified CL AGN is increasing in numbers due to dedicated search for such objects \citep[e.g.][]{macleod2019,sheng2019}. Roughly about 100 CL AGN are known so far. The timescales range from months to years, even in rather massive quasars.  

The whole group does not have to be a homogeneous sample, and perhaps different mechanisms can operate in different sources, depending whether we a brightening effect, dimming effect or quasi periodic changes. The list of candidate processes is long. In some sources the change can be actually caused by obscuration, in other sources there are strong arguments for intrinsic changes. For example, the dramatic weakening of the lines in NGC 5548 was accompanied by a decrease of the BLR size by large factor, clearly in response to the decreased irradiation flux. Such intrinsic changes can be caused by tidal disruption event, but this would be unlikely if the episodes repeat. In such case we can have orbiting clouds or stars perturbing the disk \citep{bon2016}. We can have changes of the disk state - temporary development and disappearance of the disk warm corona caused by a modification of the acccretion rate modifies the bolometric luminosity of the source but, more importantly, changes the spectral shape of the radiation flux emitted by the nucleus \citep{noda2018}. \citet{sniegowska2019} proposed that a combination of the inner hot flow with outer standard disk can provide the requested timescales for periodic changes if the radiation pressure instability operates only in a narrow disk zone which shortens the viscous timescale considerably.

This rapidly developing topic offers a new window to study BLR since instead of a semi-static picture (apart from the response of the BLR to the variable irradiating flux) we are offered a time-resolved picture of the BLR formation which will put strong constraints on the actual BLR mechanism. However, to achieve that, first some classification studies have to be done, to make sure what is the role of the obscuration in the process, whether indeed a fresh material arrived as in case of tidal disruption events. Then we have to carefully approach the issue whether the BLR material appears/disappears, or just a variable irradiation activates/disactivates the emission line production. With hundreds of objects expected soon, this will bring a very important development for the understanding how BLR works, bit also how the accretion pattern in direct vicinity of the black hole can rapidly change in some (which ?) circumstances. But it will also push the research further into time domain, since the physical formation/disappearance of the BLR takes time, and then the system contains a memory of the past state. Simple AGN/BLR unification schemes of the past do not contain this element. 

\section{Conclusions and future prospects}
\label{sec:conc}

Our understanding of the BLR formation is still very limited, but the observational progress is considerable, and accelerating. The basic properties - relatively flat distribution of emitting material, approximately in Keplerian motion are confirmed. Some of the models bring attractive explanation of the inner radius of the BLR \citep{ch2011}, outer radius \citep{netzer1993}, and local cloud density \citep{baskin2018}, but the BLR/accretion disk relation remains to be firmly established.

We still need much better understanding of the dynamics of the material, its origin, and its properties, including precise determination of the metallicity. But there is a revived determination to achieve this goal. This is related to understanding that BLR is an excellent tracer of a number of processes important for the galaxy evolution: (i) outflow and feedback (ii) stellar disruptions (iii) binary black holes (iv) drastic changes in the accretion pattern close to a black hole (v) past activity of a nucleus. Finally, prospects to use BLR for cosmology are particularly attractive \citep[see e.g.][for a review]{czerny2018}, with some quasar-based results pointing toward tension with standard $\Lambda$CDM cosmology \citep{lusso2019}.

To meet these expectations, massive spectroscopic surveys are planned to performed or already under way. SDSS-RM mapping program continues\footnote{https://www.sdss.org/dr14/algorithms/ancillary/eboss/}. The oz-DES monitoring project for a $\sim 500$ quasars started six years ago \citep{king2015} should bring soon planned measurements.  Combined effort of a number of dedicated monitoring programs \citep[e.g.][]{lira2018,woo2019,pancoast2019,dupu2018,huang2019} also continuously bring more data. Existing photometric surveys like Zwicky Transient Facility\footnote{https://www.ztf.caltech.edu/} bring notification of rapid changes in activity of galaxies, allowing for spectroscopic follow-up of interesting events. Next major step will come with SDSS-V  which will be the first facility providing multi-epoch, all-sky, optical and IR spectroscopy. With two sites, it will bring monitoring information about several thousands of supermassive black hole behaviour across the whole sky. Finally, Large Synoptic Survey Telescope (LSST)\footnote{https://www.lsst.org/} will bring numeous photometric lightcurves in six bands from the basic survey as well as from the Dip Drilling Fields which can be used for time delay measurements.



\section*{Acknowledgements}

I am grateful to the two anonymous referees for very helpful comments which lead to expansion of the manuscript. The project was partially supported by National Science Centre, Poland, grant No. 2017/26/A/ST9/00756 (Maestro 9), and by the grant MNiSW grant DIR/WK/2018/12. This work was performed in part
at Aspen Center for Physics, which is supported by National Science Foundation grant PHY-1607611. This work was
partially supported by a grant from the Simons Foundation.

\bibliography{BLR_Serbia}
\bibliographystyle{aasjournal}

\end{document}